# Learning by Test-infecting Symmetric Ciphers

K. S. Ooi

School of Science and Mathematics
Inti College Malaysia
Jalan BBN 12/1, Bandar Baru Nilai
71800 Negeri sembilan, Malaysia

Jan 8, 2006

**Abstract**

We describe a novel way in which students can learn the cipher systems without much supervision. In this work we focus on learning symmetric ciphers by altering them using the agile development approach. Two agile approaches the e*X*treme *P*rogramming (XP) and the closely related *T*est-*D*riven *D*evelopment (TDD) are mentioned or discussed. To facilitate this development we experiment with an approach that is based on refactoring, with *JUnit* serves as the automatic testing framework. In this work we exemplify our learning approach by test-infecting the Vernam cipher, an aged but still widely used stream cipher. One can replace the cipher with another symmetric cipher with the same behavior. Software testing is briefly described. Just-in-time introduction to Object-oriented programming (OOP), exemplified by using Java$^{\text{TM}}$, is advocated. Refactoring exercises, as argued, are kept strategically simple so that they do not become intensive class redesign exercises. The use of free or open-source tools and frameworks is mentioned.

## Introduction

Cryptography is only one component of the chain of components of a security system. It is nevertheless the most important component that must be done right at the beginning at all cost, because when it were broken the whole security system can be bypassed and the system that it is supposed to protect can be seemingly entered legitimately. Serious students in cryptography, in particular, and security, in general, must acquire a sound knowledge of the ciphers used in security systems. Learning the inner workings of cipher systems is thus the central issue addressed by this work. One of many ways a novice cryptologist may undertake to learn the inner workings of a cryptosystem is by attacking them [1, 2]. The essential prerequisite in this kind of study is, however, a sound, preferably thorough, understanding of the system. This work describes a scheme how you, the instructor, can help students to get closer to achieving that thorough understanding.

Cryptography is a wide and varied field that attracts lots of young, aspiring scientists, engineers, and mathematicians. New comers to this study are destined to face insurmountable difficulties or challenges. There are always new things to learn. Students must pace the learning at the rate that one could hardly imagine in pursuing any other fields of research. Let's quote Ferguson and Schneier:

> *It is impossible to understand it all. There is nobody in the world who knows everything about cryptography. There isn't even anybody who knows most of it. We* *certainly don't know everything there is to know about the subject of this book.* [3]

I always wanted to help students to learn cryptography by being their personal tutors and by treating the creation of crypto systems as crafts, as inspired by the work of McBreen [4]. Of course, this is particularly far from impossible if you were working in a teaching college. I did not give up easily without taking up the challenge to make the students' learning more personal. After working with students for some time, I devised a novel way to help them learn cryptography by themselves and, to some extend, by their own pace; that is, I test-infected a few ciphers covered in my lecture, gave them a number of refactoring exercises that are easy enough to carry out, and encourage the students to perform pair programming, which is an essential practice of a software engineering process model called e*X*treme *P*rogramming (XP). Despite being more like a practical programming exercises than mathematical ones (as many lectures in cryptography usually assume), I found that students are generally have better and firmer grounding than my previous students who had not been exposed to this learning experience. Moreover, students who further their studies in mathematical cryptography have a formal encounter or experience with practical cryptography, one aspect of *everything theres to know about the subject*. Students who may want to pursue a practical career have now a wider perspective about what to expect in their future.

A successful scientific undertaking is usually a collaborative effort. Students should be made aware of this fact. I enlisted the help of XP because I have read about the claim made by Kent Beck:

> *XP is my attempt to reconcile humanity and productivity in my own practice of software development and to share that reconciliation. I had begun to notice that the more humanely I treated myself and others, the more productive we all became. The key to success lies not in self-mortification but in acceptance that we are people in a person-to-person business.* [5]

One of the primary practices in XP is pair programming [6]. Among the benefits of pair programming are keeping each other on track, clarifying ideas, brainstorming refinements to specifications and so forth. The success stories of the approach, though you may have said that XP is nothing more than good working habits tagged with a name, is why I promoted it to the students. I can effortlessly convince the students to collaborate rather than working in seclusion. I did not, however, enforce or practice pure XP, which in its purer form has values, practices, activities and roles. You will see that my approach is rather agile, which generally means lightweight processes. Students were encouraged to read the standard work and try some practices on their own initiatives. The standard work by Kent Beck [5] is a well-written book; you could distinctly hear students talking about it passionately in lab sessions not long after it was made an additional but nonessential reading. No cryptography student should miss reading a book that contains a passage like this: "*XP also encourages human contact among the team, reducing the loneliness that is often at the heart of job dissatisfaction.*"[5].

No mention will be made about installing software packages or tools, open source or otherwise, in this work. You have to visit the related websites for details. Nor will I touch on how to compile and run programs under these environments. For the sake of completeness, in the next section I will outline the principles of software testing, an integral part of software engineering process. No explicit mention will be made to the benefits of software testing or how it should be done right, as this is beyond the scope of this work. Emphasis will be placed on unit testing – which is the heart of this work. Fuller account can be obtained from the common and standard literature that is voluminous and can be dry [8, 9], as testing is a *demanding* and a *costly* activity.

# Software Testing – A Brief Discourse

Software testing, which is originally meant to discover and reveal software errors [7] by executing the code, is a viable investment in software projects, from large to small. This is an established field whose literature runs into thousands or more. There are countless ways of saying the same things in this field. I will stick to a set of definitions that I feel comfortable with.

Software errors generally fall into one of the three categories: wrong, missing, and extra [9]. Only the wrong is relevant, of which we are able to reveal by means of testing. The collective term verification and validation, V&V in short, is a fairly common software engineering strategy to guard against the wrong. The members of this term have different meanings. It is best to distinguish them by asking questions [10]: Verification answers the question "Are we building the product right?" and validation answers the question "Are we building the right product?". Testing is thus a validation process against specifications. Functional testing, also known as black box testing, is used to develop test cases in this article. Structural testing, also known as white box testing, will be used primarily for verification.

Tests not only will lead to validation, it also leads to defect testing – which in the context of this work is of utmost relevant. Defect testing will be the key to our novel way that helps students learn the cipher systems. There are four types of general tests: unit testing, integrated testing, system testing, and user acceptance testing [9]. The former two tests are more relevant to our discourse, particularly unit testing, which ideally should be the central testing task of the programmer who produced or refactored the unit. The term unit testing is inherited from testing procedural programs. Object-oriented programs have since brought new problems that are nonexistence in procedural programs. Object-oriented tests often speak of component testing instead of unit testing; sometimes the OO practitioners confuse about unit testing and integrated testing. In this work this trivial issue does not arise because automated tests have already produced for the students. Paying much attention to this issue will only be a major distraction from the current intent. There are still other issues not to be mentioned here, you are referred to these works [8, 9] to get a more complete picture. The notion of *test coverage* propels us to confront the question: What is covered by our unit tests? In this work we will only test against a specification (which I will describe later): so, ours is a specification-based coverage, not code-based coverage. In short, Software testing for us is a straightforward affair.

# Vernam with Object Orientation

I based the cryptography course on a textbook by Morin [11]. Vernam cipher is used in this work due to its simplicity. When the length of the key is at least that of the plaintext and the key is used only once, you have a theoretically unbreakable one-time pad. That is why Vernam cipher is still in use today for military and diplomatic communications when security is of utmost important. The basic unit of the enciphering and deciphering of the Vernam stream cipher should be a bit; we modify this requirement because it is easier to deal with bytes. Enciphering $E$ and deciphering $D$ are both accomplished by the binary operation modulo-2 addition $\oplus$ (or bitwise xor operation):

$$E(k_j, m_j) = k_j \oplus m_j = c_j$$
$$D(k_j, c_j) = k_j \oplus c_j = m_j$$

where $j$ is used as index of the $(j+1)$-th byte of a *bytestring* message $m_0 m_1 \cdots m_n \in M$, or a bytestring ciphertext $c_0 c_1 \cdots c_n \in C$, or a bytestring key $k_0 k_1 \cdots k_n \in K$. Using the open-source

CASE tool called ArgoUML [12] I produced the following class diagram with the UML notation:

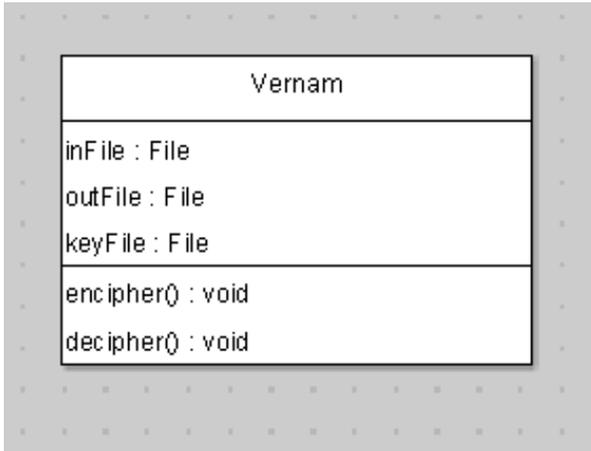

The diagram shows a simple requirement capture: an instance of the Vernam class must read in three files (`inFile`, `outFile`, and `keyFile`) and then ought to perform the operations `encipher` and `decipher`. The students should not have much problem in understanding the diagram.

Teaching object-oriented programming (OOP) to crypto students is a challenge of its own. From the start, I knew I had little time to cover OOP to the depth that the students had enough knowledge to claim equivalent credits of an undergraduate course. With a mere half an hour briefing time allocated for each two-hours-per-week computer lab session, I must be rather brief but *inspiring*. With students working in pairs the job of inspiring might be delegated to the alpha, or occasionally the beta, of the pair. The introduction of the subject is based on general principles of OOP rather than specifically slanted to one particular programming language. Java$^{\text{TM}}$ was selected based on its availability in our computer labs. The `Vernam` class that is deliberately coded with duplication in Java$^{\text{TM}}$ is shown as follows:

```java
import java.io.*;
import java.util.*;
/**
 *  This is the original version. It has duplication
 *  yet to be eliminated.
 */
public class Vernam{
   // Attributes
   private File inFile;
   private File outFile;
   private File keyFile;
   // Constructor
   public Vernam(File inFile, File outFile, File keyFile){
      this.inFile  = inFile;
      this.outFile = outFile;
      this.keyFile = keyFile;
   }
   // enciphering method
   public void encipher() throws FileNotFoundException, IOException{
      FileInputStream  fin  = new FileInputStream(inFile);
```

```
        FileInputStream  fkey = new FileInputStream(keyFile);
        FileOutputStream fout = new FileOutputStream(outFile);
        int m = -1, k = -1;
        while( (m = fin.read() ) != -1 ){
            k = fkey.read();
            fout.write(k^m);
        }
        fin.close();
        fout.close();
        fkey.close();
    }
    // deciphering method
    public void decipher() throws FileNotFoundException, IOException{
        FileInputStream  fin  = new FileInputStream(inFile);
        FileInputStream  fkey = new FileInputStream(keyFile);
        FileOutputStream fout = new FileOutputStream(outFile);
        int m = -1, k = -1;
        while( (m = fin.read() ) != -1 ){
            k = fkey.read();
            fout.write(k^m);
        }
        fin.close();
        fout.close();
        fkey.close();
    }
}
```

---

Since no knowledge in Java$^{\text{TM}}$ programming language was presumed, I breezed through the followings:

- What a class is about?
- Keywords and identifiers.
- Native and abstract data types.
- Constructors of a class.
- Access modifiers.
- Class methods and their invocation.

In the next two paragraphs I demonstrate how I taught all these just-in-time!

The `Vernam` class is an abstraction of the Vernam stream cipher. You may think of the class as a set of objects with similar behavior. We will see later how to create an object using the class as a factory. We use object and instance interchangeably – they both mean an individual representative of a class with specified attributes. The behavior of an object is about what it can perform. An object of `Vernam` can perform enciphering and deciphering, and that defines its behavior. The attribute of an object is all the information held by it at a given time. If we specify the `inFile`, `outFile`, and `keyFile` of a `Vernam` object, we define its attributes.

Java<sup>TM</sup>, aside being a programming language, can be viewed as a framework that contains many packages by which we can use to create practical applications. We use the keyword `import` to enable us to use a specific package from the framework. The keyword appears twice in the class. Packages are grouped. Package names can be omitted using *. Seven keywords are used in the class: `public`, `class`, `private`, `this`, `void`, `throws`, `new`, and `int`. These are reserved words of the language, which mean you should not use them as variable names. Variable names or identifiers appear in the class are `inFile`, `outFile`, `keyFile`, `fin`, `fkey`, `fout`, `m`, and `k`. Only one primitive type, that is type `int`, is used to define `m` and `k`. The rest are all class types, including `File`, `FileInputStream`, and `FileOutputStream`.

The class contains a non-default constructor, which has the same name as the class. It is used to create new `Vernam` objects in the client codes. The constructor does not have a return type. All attributes are `private`, which means they cannot be accessed outside the class. Therefore the class is designed in the way that you can *set these fields* of an object during its creation by passing the three parameters to the constructor. Inside the body of the constructor, `this` keyword is used to resolve the names of the parameters and the attributes. This constructor is made `public`, so are the two class methods `encipher` and `decipher`. We construct a new object using the `new` operator, as exemplified by the creation of a `FileInputStream` object identified by the identifier `fin` in the class method `encipher` using the constructor `FileInputStream(inFile)`. The `encipher` method has a `while` loop. The condition of the loop combines an input statement that assigns the byte read to the variable `m` and a test for the end of the input file `fin`. This code is rather messy, but it is widely used in this way. Also we invoke a `read` method *on* `fin`. The modulo-2 addition is translated in Java<sup>TM</sup> as the bitwise xor operator $^\wedge$. The `decipher` method is a duplication by deliberation.

The preceding two paragraphs are all that is 'necessary' to cover OOP just-in-time. OOP should be targeted at the class produced; I do not recommend teaching a general OOP to cryptography students from the beginning. Java<sup>TM</sup> IO should not be covered at this stage, because it can be a course of its own – what the students must be aware of is that the `encipher` and `decipher` methods read the input file and the key file byte-by-byte and write the result of the modulo-2 addition to an output file.

Think of an object-oriented system as a community that contains interacting objects. Each object provides a service that is used by others so that a meaningful application can be constructed. In the class methods of `Vernam` you can see how this view takes shape. I then wrote a client class to create an enciphering `Vernam` object identified by `en` and invoke the `encipher` method on it. A deciphering object called `de` was also created. The client class is as follows:

```
import java.io.*;
public class RunVernam{
   // the main method
   public static void main(String[] args){
      try{
         Vernam en = new Vernam(new File("in.jpg"),
                   new File("enciphered.jpg"), new File("key.jpg"));
         en.encipher();
         Vernam de = new Vernam(new File("en.jpg"),
                   new File("deciphered.jpg"), new File("key.jpg"));
         de.decipher();
      }
```

```
        catch(FileNotFoundException e){
            e.printStackTrace();
        }
        catch(IOException e){
            e.printStackTrace();
        }
    }
}
```

---

The client class `RunVernam` comes with the `main` method, starting from which the program will be executed. The parameter that holds the command-line arguments is not used. Since the `encipher` method and the `decipher` method both throw `FileNotFoundException` and `IOException`, the `try-catch` blocks must be in place to catch them. A successful Vernam cipher should yield the same persistent files `in.jpg` and `deciphered.jpg` (You may want to use any files on your discretion). Our specification-based coverage is thus as follows:

$$m_j = D(k_j, E(k_j, m_j))$$

We will test against this specification.

## Automated Tests with JUnit

In this work we use the open source framework *JUnit* [13] originally written by Kent Beck and Erich Gamma. JUnit *is becoming* the de facto standard tool for Java$^{\text{TM}}$ unit testing. You may want to read the article by its originators [15]. In writing the test program, which I rightfully call `TestVernam`, I extend the class from the `TestCase`. Two things worth your attention in the class: first, the `isEqual` method that checks the equality of two persistent files byte-by-byte; second, the `assertTrue` method from `TestCase` whose condition tests our specification-based coverage. The code from the `RunVernam` has been moved to this class. We no longer need to run `RunVernam`. JUnit is, of course, more than just `TestCase`. We are only concerned with `TestCase` at this moment. There are many interesting packages that you might find suitable to other aspects of unit testing. The `TestVernam` class is as follows:

---

```
import junit.framework.TestCase;
import java.io.*;
public class TestVernam extends TestCase{
  // default constructor
  public TestVernam(){
    super();
    setFiles(new File("in.jpg"),new File("key.jpg"),
        new File("encrypted.jpg"),new File("decrypted.jpg"));
  }
  // set attributes method
  public void setFiles(File inFile, File keyFile, File encrypted,
                       File decrypted){
    this.inFile   = inFile;
    this.keyFile = keyFile;
    this.encrypted = encrypted;
```

```java
      this.decrypted = decrypted;
      if(encrypted.exists())
         encrypted.delete();
      if(decrypted.exists())
         decrypted.delete();
   }
   // method that checks whether file1 == file2
   public int isEqual(File file1, File file2)
      throws FileNotFoundException, IOException
   {
      if(file1.length() != file2.length())
         return -1;
      FileInputStream s1  = new FileInputStream(file1);
      FileInputStream s2  = new FileInputStream(file2);
      int n1 = -1, n2 = -2;
      while( (n1 = s1.read() ) != -1){
         n2 = s2.read();
         if(n1 != n2)
            return -1;
      }
      s1.close(); s1 = null;
      s2.close(); s2 = null;
      return 0;
   }
   // testVernam method
   public void testVernam()
   {
      try{
         Vernam en = new Vernam(inFile,encrypted,keyFile);
         en.encipher();
         Vernam de = new Vernam(encrypted,decrypted,keyFile);
         de.decipher();
         assertTrue( 0 == isEqual(inFile,decrypted) );
      }
      catch(FileNotFoundException e){
         e.printStackTrace();
      }
      catch(IOException e){
         e.printStackTrace();
      }
   }
   // private attributes
   private File inFile    = null;
   private File keyFile   = null;
   private File encrypted = null;
   private File decrypted = null;
}
```

Students need not know much about the class, since it has been produced for them. However,

students who want to recreate the `Vernam` class from scratch that defines new message-passing protocols, whereby the behavior of its objects has changed, and/or different coverage must study the class in details. All Student pairs must be able to run the *TestRunner* on `TestVernam` by themselves in lab sessions. The successful test produces the following graphical (only the Swing based TestRunner is shown here), somewhat joyful, announcement:

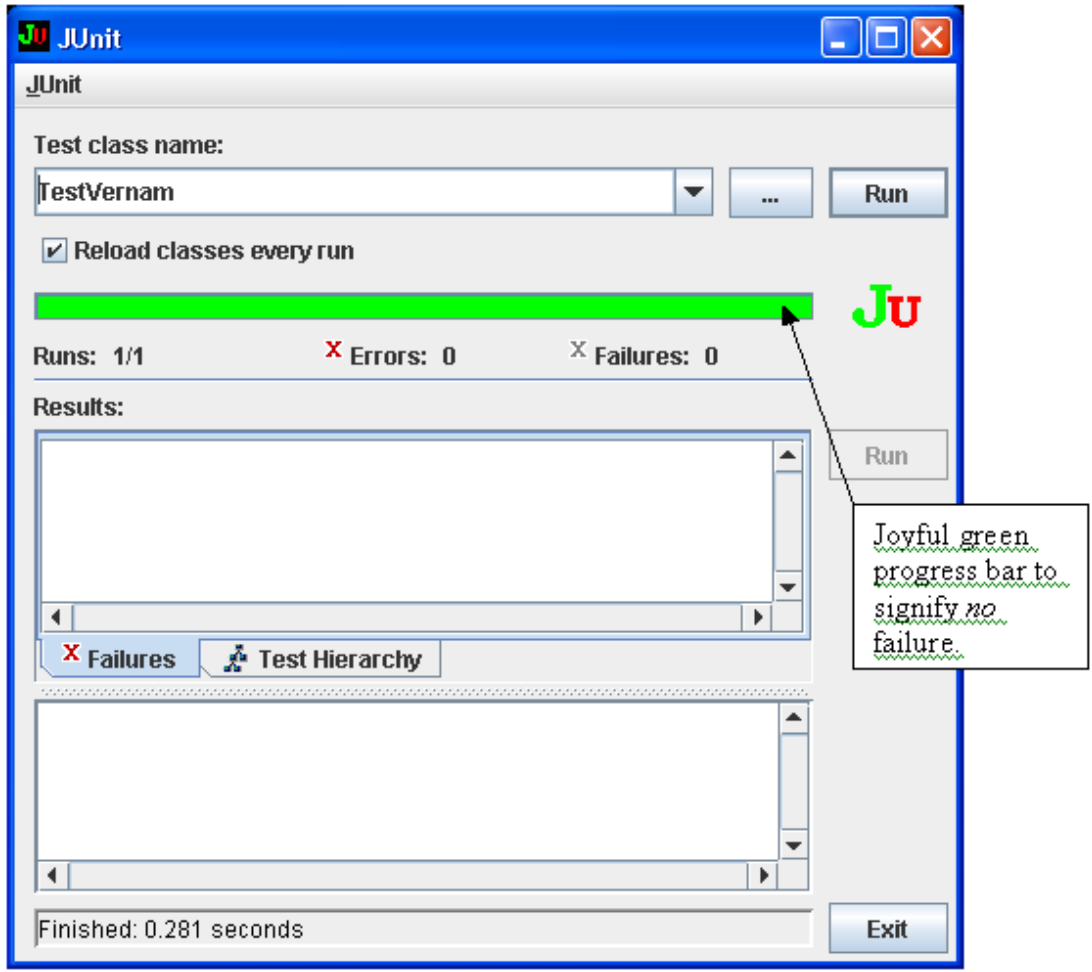

If I changed the bitwise xor $^\wedge$ into bitwise or | in the `decipher` method of `Vernam` in an fault-based testing exercise and then compiled, and run the TestRunner on `TestVernam`, an eerie red progress bar announces that the test has failed:

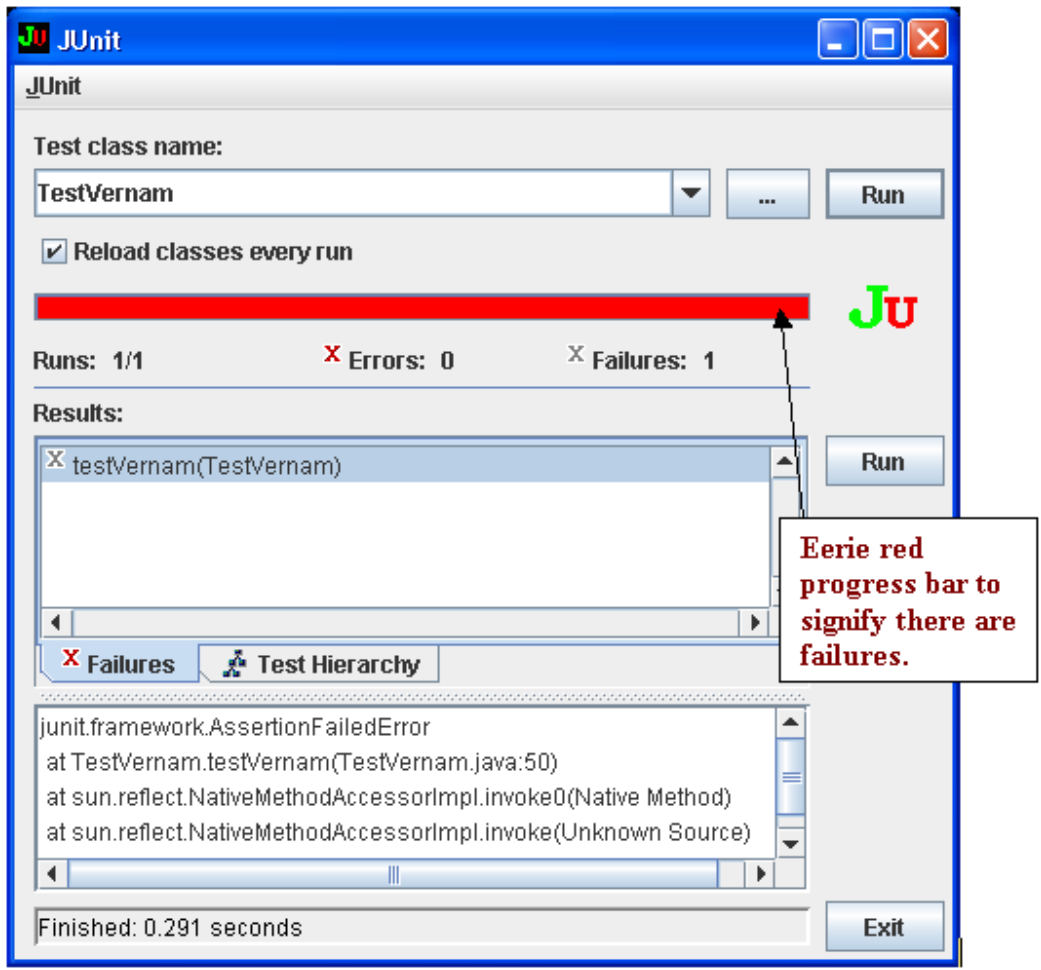

The main advantage of running the TestRunner on `TestVernam` rather than running the `RunVernam` class is that, after you have made changes to the `Vernam` class, the test is just one click away. This fact is clearly illustrated in the next section. The testing seems easy to carry out once the `TestVernam` class is produced, because it is automated.

## Refactoring Exercises

The refactoring exercises can be accomplished according to the following test loop:

1. The pair programmers run the TestRunner on `TestVernam`
2. They make changes to the `Vernam` class by refactoring
3. They compile the `Vernam` class *successfully*
4. They click the Run button on the TestRunner
5. They repeat step 2 if there are other refactoring exercises to be considered.

You get a better picture how the test-loop works in the following UML activity diagram, which was also produced using ArgoUML:

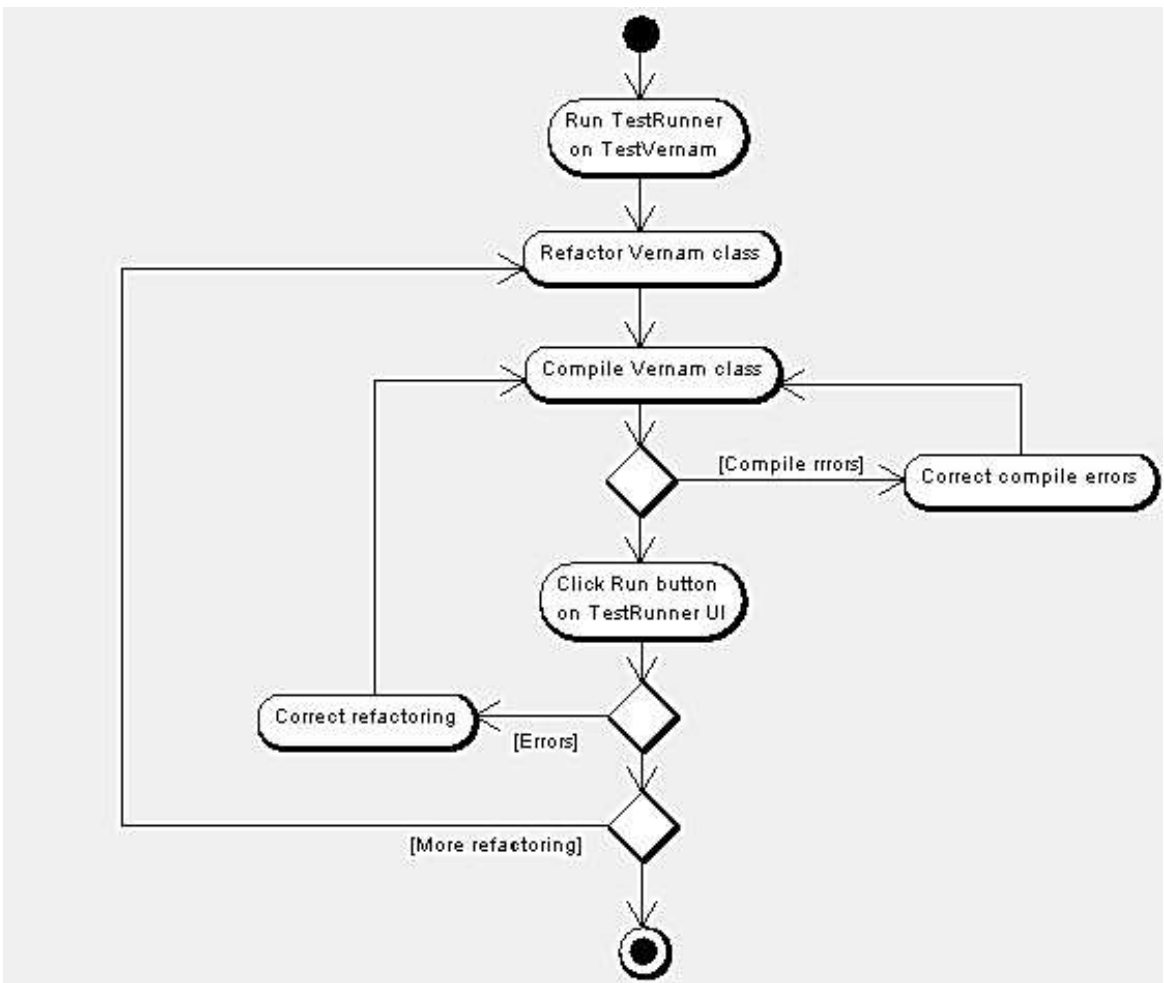

The `Vernam` class is far from ideal. One obvious 'eyesore' is the duplication codes in encipher and decipher class methods. Getting rid of duplication is a must, because it can be easily accomplished by students who have had no prior knowledge in programming. Students were prompted to create a common method called `xor` that holds the common codes. This is how the refactoring loop gets underway in most of the lab sessions. If a student wanted to *recreate* the whole `Vernam` class, to whatever extent, they can do so by modifying and compiling the `TestVernam` class. If the behavior of the cipher objects remains the same, recreating the cipher will be much more simpler.

The size of the key file was never checked in the `Vernam` class – this has infringed the one-time pad requirements. Throwing an exception is a conventional way to handle an error in Java$^{TM}$. I recommended that the students should insert the following code snippet that throws an unchecked exception into the class, and then run a fault-based testing exercise.

```
if(keyFile.length() < inFile.length())
    throw new IllegalArgumentException("Key file < the input file.");
```

These are the two main, simple refactoring exercises that students of cryptography are able to accomplish within a week or two. You can come up with more exercises; there are many possibilities. New and 'better' technologies are constantly added to the Java$^{TM}$ framework. For example you might want to refactor the `IO` to `NIO`. You, as an instructor, will be the only person who is able to decide the appropriateness of the new refactoring exercises. The standard work on refactoring by Fowler [14] should be consulted cautiously so that refactoring exercises does not become redesign exercises.

# Discussion and Conclusion

The choice of Vernam cipher is appropriate. It is a cipher simple enough that it does not muddle the main intent of this work. You might as well replace it with another block cipher of your choice in your refactoring exercises. I have achieved similar result with AES based on yet another testing model [16]. During the refactoring process, students were encouraged to perform verification, which involved activities like code inspection, code walkthrough, and design review. Some may prefer to introduce interface class to enforce integrity of protocol. Java$^{\text{TM}}$, however, was not explicitly taught in lab sessions. A very brief introduction based on the class was delivered just-in-time. If I replaced the block cipher with another one, I would 'rehearse' OOP again in the context of the new class. With pair programming the students seem to learn it by self-motivation. All the activities (that include our test-infecting process, verification, and pair programming) do seem to improve the understanding of the cipher systems that undergo the test loop. If you were the instructor of a cryptography course, you might want to try out one or two lab sessions to see the results. The harmful effect is really minimal.

Cryptography course can be taught without computers or lab sessions. Nevertheless, it will only benefit a few students who are 'talented' in mathematics; in community colleges these students are indeed rare. A broader range of students might be able to enjoy success from cryptography if they were given ample opportunities to try out things by themselves with actual number of bits. They will only be able to play with toy numbers in most circumstances without computers. For example, a recent textbook on cryptography by Trappe and Lawrence [19], which is based on courses taught at the University of Maryland and Rutgers University, contains computers examples written in Mathematica®, Maple® and MATLAB®. The book encourages instructors and students alike to use computer programs in conjunction with cryptography courses. The programming languages used, though user friendly, are lack of flexibilities, consistencies and advanced features found in object-oriented programming languages like Java$^{\text{TM}}$ or C# that comes with a well-tested framework. Even though I use Java$^{\text{TM}}$ in my lab sessions, the use of C# does not seem to cost lots of uneasiness among the students [16].

Another related development approach, the test-driven development (TDD), whose goal is to produce *clean code that works* [18], was briefly described to the students. TDD is considered as *an awareness of the gap between decision and feedback during programming and techniques to control that gap* [18], which may be deemed as the less absolute XP. Using test-driven development to learn Java$^{\text{TM}}$ has been recently championed in the book by Langr [17]. In TDD, the development is driven by writing automated tests *first*. Thus shorter feedback loops are assured in development. Students who are interested in more programming tasks are encouraged to try out this approach.

In my lab sessions I had said: "We know that any nontrivial software has errors/bugs. It is inevitable. Passing various tests is not enough to guarantee the absence of errors, unless we are able to test all combinatorial inputs, which is impossible as the number of test cases is astronomical. Hence you should not use testing to prove a system. But we could easily use testing to enhance our understanding about the inner working of cipher systems." There will always be a couple of skeptical students who feel that testing should be as good as proving. These students can then be asked to develop test cases for a really trivial program that reads three integers as the sides of a triangle and print a message that states whether the triangle is scalene, isosceles or equilateral (posed by Myers in 1978) [20]. There are so many resources that can help you. You have little worries about failure using the learning system I have described. It might work so well with students.

Our test-infected (that is the *test-infecting* of the title of this work) 'fever' is borrowed from Beck and Gamma [15]. Discussion on fault-directed testing, one other way to boost confidence on a cipher system in addition to learning it, in spite of its relevance, will be delayed to another work [16]. To conclude, the test loop *is* able to complete or complement the mathematical introduction to Vernam stream cipher from the lecture.

## Short Autobiographical Sketch

Kuan-San Ooi obtained a PhD degree from the University of Malaya, Malaysia, in 1997, specializing in statistical thermodynamics and scientific computing. Prior to joining Inti College Malaysia as a senior lecturer, he founded the crypto lab and was a senior lecturer at the University of Sheffield Center at Taylor's College, Malaysia. He is interested in cryptography and human learning research.

## Acknowledgements

I must thank Mr. K. E. Foo for his interest in the project and his efforts to LaTeX the document. I thank Dr. S. C. Soh for identifying some errors in the early draft.

## References

[1]   Bruce Scheier, *Self-Study Course in Block Cipher Cryptanalysis*, Cryptologia, 24(1), 18-34, 2000

[2]   K. S. Ooi and Brain Chin Vito, *Cryptanalysis of S-DES*, Cryptology ePrint Archive: Report 2002/045, 2002

[3]   Niels Ferguson and Bruce Schneier, *Practical Cryptography*, Wiley Publishing, Inc., Indianapolis, Indiana, 2003

[4]   Pete McBreen, *Software Craftsmanship: The New Imperative*, Addison-Wesley Professional, 2001

[5]   Kent Beck and Cynthia Andres, *eXtreme Programming Explained: Embrace Change*, 2nd Edition, Addison Wesley Professionals, 2004

[6]   Laurie Williams and Robert Kessler, *Pair Programming Illustrated*, Addison-Wesley, 2002

[7]   Edsger W. Dijkstra, *On the cruelty of Really Teaching Computer Science*, Communications of the ACM, 32(12), 1398-1404, 1989

[8]   Robert V. Binder, *Testing Object-Oriented Systems: Models, Patterns, and Tools*, Addison-Wesley, Reading, Massachusetts, 2000

[9]   William E. Perry, *Effective Methods for Software Testing*, 2nd Edition, John Wiley & Sons, New York, 2000

[10]  Barry W. Boehm, *Software Engineering: R & D trends and defense needs, In Research Directions in Software Technology, Peter Wegner (Ed.)*, MIT Press, Cambridge, MA, 1979

[11] Richard A. Morin, *An Introduction to Cryptography*, Chapman & Hall/CRC, Boca Raton, Florida, 2001

[12] ArgoUML project at *http://argouml.tigris.org*

[13] JUnit project at *http://www.junit.org*

[14] Martin Fowler, *Refactoring: Improving the Design of Existing Code*, Addison-Wesley, Boston, 2000

[15] Kent Beck and Erich Gamma, *Test-Infected: Programmers Love Writing Tests*, JavaReport, July 1998

[16] K. S. Ooi, *Test-Infecting AES*, to be published.

[17] Jeff Langr, *Agile Java: Crafting Code with Test-Driven Development*, Prentice-Hall PTR, 2005

[18] Kent Beck, *Test-Driven Development by Examples*, Addison-Wesley, 2002.

[19] Wade Trapple and Lawrence Washington, *Introduction to Cryptography with Coding Theory*, 2nd Edition, Pearson Education, 2006

[20] Glenford J. Myers, *The Art of Software Testing*, John Wiley & Sons, New York, 1979